\documentclass[12pt]{article}
\usepackage{microtype}
\usepackage{amssymb}
\usepackage{ragged2e}
\usepackage{graphicx}
\usepackage{xcolor}
\usepackage{amsmath}
\usepackage{booktabs}
\usepackage[absolute,overlay]{textpos}
\usepackage{appendix}
\usepackage{verbatim}
\usepackage{mathrsfs}
\usepackage{pdfpages}
\usepackage{capt-of} 
\usepackage{caption}
\usepackage[normalem]{ulem}
\usepackage{float}     
\usepackage{subfig}    
\usepackage{placeins}
\usepackage{longtable}
\usepackage{amsmath}
\usepackage{array}
\usepackage{multirow}
\usepackage{authblk}
\usepackage[margin=1in]{geometry}
\usepackage{nameref}
\usepackage{indentfirst} 
\usepackage[normalem]{ulem} 

\usepackage{hyperref}

\title{{Resonances in the early Universe}}
\author[1 \footnote{daniel.canedo@estudante.ufjf.br }]{D. L. Canedo}
\author[1 \footnote{gilnetojf@ufjf.br}]{G. Oliveira-Neto}
\author[2 \footnote{monerat@iprj.uerj.br}]{G. A. Monerat}
\author[3 \footnote{profeduvasquezuerj@gmail.com.br}]{E.V. Corr\^{e}a Silva}

\affil[1]{Departamento de Física, Instituto de Ciências Exatas, Universidade Federal de Juiz de Fora, Juiz de Fora--MG, CEP 36036-330, Brazil}
\affil[2]{Departamento de Modelagem Computacional, 
		Universidade do Estado do Rio de Janeiro, Instituto Polit\'{e}cnico, 
		Rua Bonfim, 25, Vila Am\'{e}lia, Nova Friburgo--RJ, CEP 28625-570, Brazil}
\affil[3]{Departamento de Matem\'{a}tica, F\'{\i}sica e Computa\c{c}\~{a}o, Faculdade de Tecnologia,
	Universidade do Estado do Rio de Janeiro, Av. Dr. Omar Dibo Calixto Afrange s/n, Resende--RJ, CEP 27537-000, Brazil}

\date{December 15, 2025}

\begin{document}

\maketitle

\begin{center}
 \section*{Abstract} \justify
In the present paper, we study a Friedmann-Lemaître-Robertson-Walker (FLRW) quantum cosmology model with positively curved spatial sections. The matter content of the model is given by a radiation fluid, a constant vacuum energy, and an ad hoc potential. After writing the Hamiltonian of the model, we notice that the effective potential ($V_{eff}$) depends on two parameters: $\rho_v$, the constant vacuum energy density and $\sigma$, associated with the ad hoc potential. Depending on the values of these parameters $V_{eff}$ becomes a double barrier potential. We quantize the model and obtain the Wheeler-DeWitt equation. We solve that equation using the WKB approximation and compute the corresponding probability ($TP_{WKB}$) that the wavefunction of the universe tunnels through the double barrier potential $V_{eff}$. We study how $TP_{WKB}$ behaves as a function of the parameters $\rho_v$, $\sigma$ and the radiation energy $E$. We notice the occurrence of resonances in $TP_{WKB}$ when we vary it as a function of $E$, $\rho_v$ and $\sigma$. It is a very interesting phenomenon because it may cause the universe to be born with selected values of $E$, $\rho_v$ and $\sigma$.
\end{center}


\section{Introduction}










The application of General Relativity (GR) in Cosmology has brought important results for the understanding of the early Universe \cite{Guth}. Over the years, several cosmological models, with different geometries and different matter contents, have been proposed in order to describe the Primordial Universe. 
Unfortunately, GR fails when we apply it to study the birth of the universe. This problem is known as the initial singularity problem \cite{Hawking1,Hawking2,Hawking3}. With the aim of studying the birth of the universe and thus eliminating the initial singularity problem, Quantum Cosmology (QC) emerged, unifying the concepts of Cosmology and Quantum Mechanics, where the Universe is described as a quantum mechanical system \cite{QC1,QC2,QC3,paulo,QC4}.

Quantization of cosmology has led to the discovery of an important equation called the Wheeler-DeWitt (WDW) equation \cite{WdW1,WdW2}.  This equation allows us to study the universe as a quantum mechanical system through a wave function that depends on the parameters of the model. The quantum mechanical description of the early universe introduced an important mechanism capable of removing the initial singularity: quantum tunneling \cite{Q creation1,Q creation2,Q creation3,Q creation4,Q creation5,Q creation6}. In this mechanism the universe is represented by a wave function which tunnels through a well-defined and finite potential barrier, from its initial nil size to a later non-zero size. This process eliminates the initial singularity because, upon tunneling, the Universe appears on the other side of the barrier with a nonzero size and free of the initial singularity. This concept has been explored in several works for different geometries and material contents \cite{Modelos QC1,Modelos QC3,Modelos QC4,Modelos QC5,Modelos QC6,Daniel,Modelos QC7}.


In Refs. \cite{Q creation1,Q creation2,Q creation3,Q creation4,Q creation5,Q creation6,Modelos QC1,Modelos QC3,Modelos QC4,Modelos QC5,Modelos QC6,Daniel,Modelos QC7}, mentioned above, the authors studied quantum mechanically the birth of the universe through the quantum tunneling mechanism. In all papers, the universe had to tunnel through a single barrier potential in order to be born. 
Here, we propose a more general model in which the universe has to tunnel through a double barrier potential to be born. The new ingredient we add to achieve this is the ad hoc potential, whose origin is believed to be geometric \cite{Modelos QC6} or due to a new fluid, as discussed in Section \ref{seção2}. Therefore, our model has a constant vacuum energy, a radiation fluid and an ad hoc potential in the matter sector. In order to compare the results obtained here with those in \cite{Daniel}, we will fix the curvature constant $k=1$.


A very interesting phenomenon occurs when one computes the tunneling probability for a double potential barrier, the phenomenon of resonance. It appears due to the multiple reflections of incident and reflected waves between the barriers \cite{Bohm,Merzbacher}. Resonance in physics has important applications in particle physics \cite{Ressonância2,Ressonância3,Ressonância33,Ressonância35,Ressonância1}, molecular physics \cite{Ressonância40,Ressonância4} and, more recently, in the problem of incident electrons in graphene-based double barrier structures \cite{Ressonância5,Ressonância6,Ressonância7,Ressonância8,Ressonância9}. In all these Refs. \cite{Ressonância2}-\cite{Ressonância9}, the authors studied the tunneling probability through an effective potential with a double barrier. The double barrier may be symmetric, if the two barriers are identical, or asymmetric, if the two barriers are different. For symmetric, retangular, double barriers, the tunneling probabilities at the resonances are equal to one. On the other hand, for asymmetric double barriers, the tunneling probabilities at the resonaces are smaller than one. In particular, the authors of Refs. \cite{Ressonância2,Ressonância1,Ressonância6,Ressonância7} compute the tunneling probabilities for symmetric and asymmetric double barriers in different physical situations. They explicitly show the differences in the tunneling probabilities for symmetric and asymmetric double barriers.
As described above, we study the tunneling probability of the wavefunction of the Universe through a double barrier. Due to the geometry and matter content of our model, we have an asymmetric double barrier. Thus, we are interested in studying the resonant behavior of a quantum cosmology model in the presence of an asymmetric double barrier. 

Thus, our work is structured as follows: In section \ref{seção2}, we obtain the effective potential and the Wheeler-DeWitt equation of the model. We use the WKB approximation to solve the WDW equation and thus obtain the tunneling probability $(TP_{WKB})$. In section \ref{seção3}, we present our numerical results for $TP_{WKB}$ as a function of different model parameters. In section \ref{conclusão}, we present our conclusions and discuss our results. 
This work introduces the phenomenon of resonance applied to quantum cosmology, thus contributing to the development and understanding of this field.


\section{WKB Tunneling Probability} \label{seção2}


In the present paper, we study a Friedmann-Lemaître-Robertson-Walker (FLRW) quantum cosmology model with positively curved spatial sections. The matter content of the model is given by a radiation fluid, a constant vacuum energy, and an ad hoc potential \cite{Modelos QC6,Daniel}. 
The ad hoc potential $V_{ah}$ is defined as
\begin{equation} 
\label{ad hoc}
    V_{ah} = - \frac{\sigma^{2} a^{4}}{(a^{3} + 1)^{2}},
\end{equation}
where $\sigma$ is a dimensionless parameter associated with the magnitude of that potential.
The {\it ad hoc} potential $V_{ah}$ Eq. (\ref{ad hoc}) can be physically motivated in two complementary ways.
First, it can be interpreted as an effective contribution from a more fundamental gravitational theory than General Relativity. In particular, in modified gravity scenarios, such as Ho\v{r}ava-Lifshitz gravity, purely geometric terms appear in the Hamiltonian that scale non-trivially with the scale factor.
The {\it ad hoc} potential $V_{ah}$ Eq. (\ref{ad hoc}) reproduces two relevant asymptotic regimes:
for $a \ll 1$, it behaves as $V_{ah} \to -\sigma^2 a^4$, mimicking a negative cosmological constant and generating a potential barrier; for $a \gg 1$ it behaves as $V_{ah} \to -\sigma^2/a^2$, corresponding to a stiff matter contribution with the opposite sign, which asymptotically suppresses the barrier.
For instance, these behaviors are present in the geometric sector of cosmological models constructed from Ho\v{r}ava-Lifshitz gravitational theory \cite{horava,bertolami,kord,gil3,gil}. It suggests that $V_{ah}$ can effectively encode the residual geometric effects of a more fundamental theory.

Alternatively, one may interpret the {\it ad hoc} potential $V_{ah}$ Eq. (\ref{ad hoc}) as coming from a new type of fluid with the following equation of state,
\begin{equation}
p_{ah} = \left(\frac{a^3-1}{a^3+1}\right)\rho_{ah}.
\label{eqofstate}
\end{equation}
where $p_{ah}$ is the pressure and $\rho_{ah}$ is the energy density of the {\it ad hoc} fluid.
By substituting the equation of state Eq. (\ref{eqofstate}) into the energy conservation equation for a homogeneous and isotropic universe,
one obtains the following energy density $\rho_{ah}$,
\begin{equation}
\rho_{ah} = - \frac{\sigma^2}{(1+a^3)^2},
\label{energydensity}
\end{equation}
where the parameter $\sigma^2$ is introduced as a redefinition of an integration constant. This energy density is the correct expression that leads to $V_{ah}$. From the equation of state, Eq. (\ref{eqofstate}), we notice that the {\it ad hoc} fluid behaves as a negative cosmological constant for very small values of $a$ ($a \ll 1$) and as a stiff matter perfect fluid, with the opposite sign, for very large values of $a$ ($a \gg 1$). Those are the same asymptotic limits calculated from the {\it ad hoc} potential $V_{ah}$ Eq. (\ref{ad hoc}).


From Ref. \cite{Daniel}, we learn that the total Hamiltonian ($H$) for a FLRW Universe, coupled to a radiation fluid, a constant vacuum energy, and an ad hoc potential, is given by
\begin{equation}
    H = -\frac{1}{12} P^{2}_{a} + P_{T} - V_{eff}, \label{H},
\end{equation}
where $V_{eff}$ is the effective potential and is defined as
\begin{equation} \label{Veff}
    V_{eff} = 3a^{2} - \rho_v a^{4} + \frac{\sigma^{2}a^{4}}{(a^{3} + 1)^{2}}.
\end{equation}
In the above equations (\ref{H}) and (\ref{Veff}), $P_{a}$ and $P_{T}$ are the canonically conjugate momenta of the variables $a$ and $T$, respectively. The variable $T$ is associated with the radiation fluid and plays the role of time in the model. It was obtained explicitly in Ref. \cite{Daniel}, with the aid of the Schutz variational formalism \cite{Schutz 1,Schutz 2}. $k$ is the curvature constant, $\rho_v$ is the constant vacuum energy density. For simplicity, we are using natural units, where $\hbar= 8\pi G = c = 1$.

\FloatBarrier

\begin{figure}[H]
    \centering
    \begin{minipage}{0.48\textwidth}
        \centering
        \includegraphics[width=\linewidth]{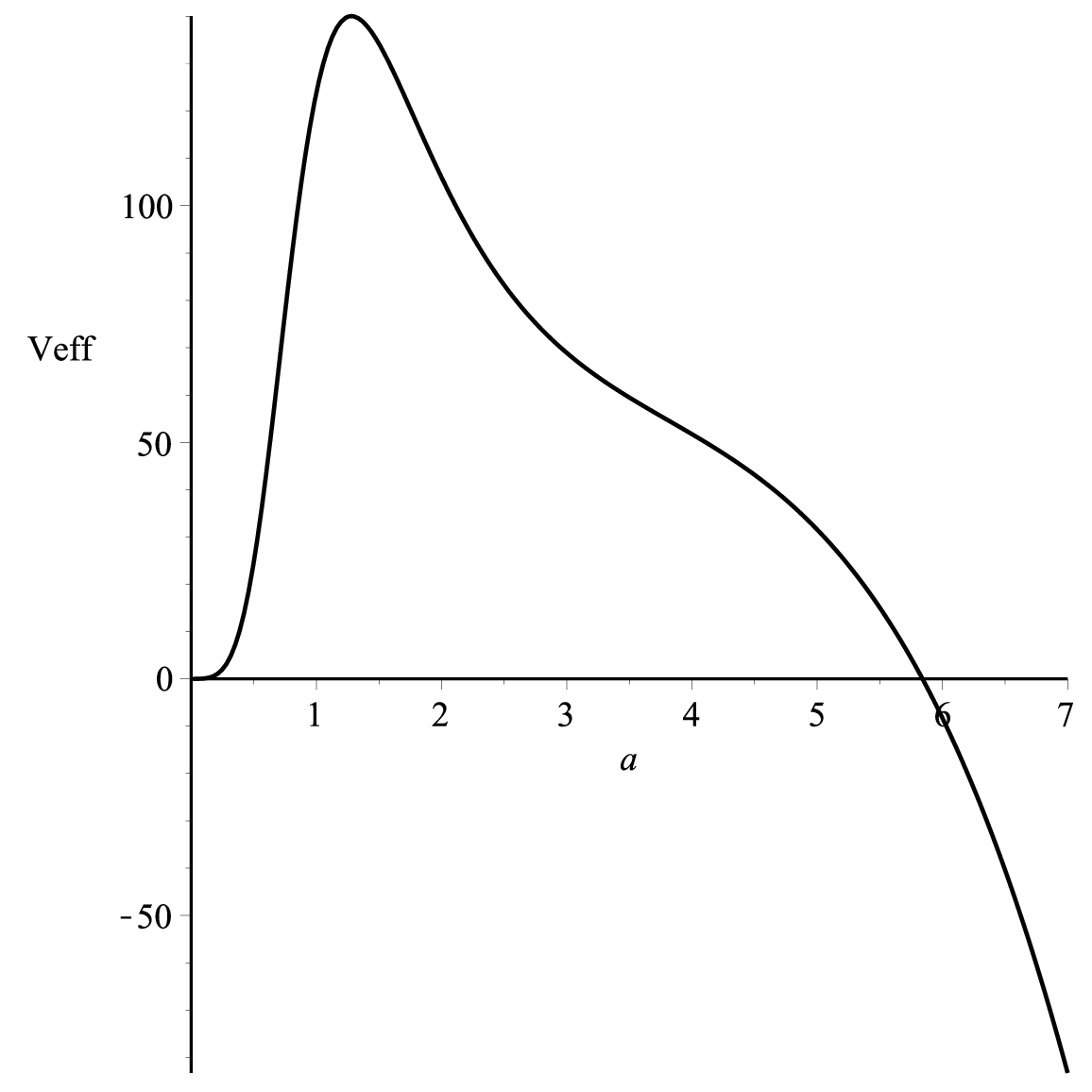}
        \caption{$V_{eff}$ (\ref{Veff}) with one barrier, where $k=1$, $\sigma=-22$, $\rho_v=0.1$.}
        \label{F1_Veff_1}
    \end{minipage}
    \hfill
    \begin{minipage}{0.48\textwidth}
        \centering
        \includegraphics[width=\linewidth]{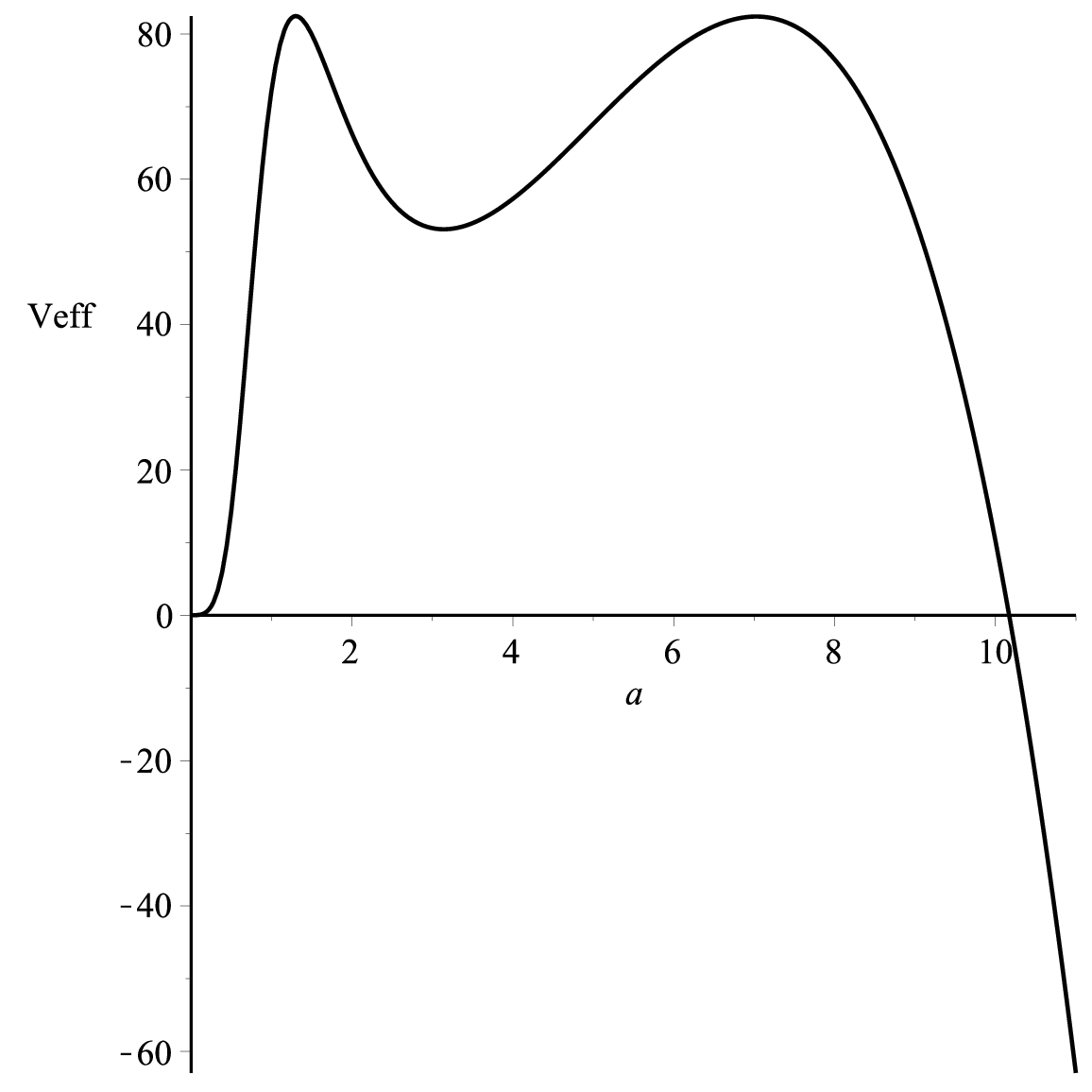}
        \caption{$V_{eff}$ (\ref{Veff}) with two barriers, where $k=1$, $\sigma=-16.65$ and $\rho_v=0.02925$.}
        \label{F1_Veff_2}
    \end{minipage}
\end{figure}

Since we are interested in comparing this study with one of the models studied in \cite{Daniel}, we fixed the constant of curvature in $k=1$. By doing so, we observe that for some values of the parameters $\rho_v$ and $\sigma$, the effective potential $V_{eff}$ Eq. (\ref{Veff}) may have one or two barriers, as we can see, respectively, in Figures \ref{F1_Veff_1} and \ref{F1_Veff_2}. In the present paper, we restrict our attention to models where $V_{eff}$ has two barriers. The models where $V_{eff}$ has only one barrier have already been investigated in Ref. \cite{Daniel}.

In order to study the probability that the Universe tunnels through the potential barriers, the quantization of the model is required. To do this, we use the Dirac formalism, where the variables, the canonically conjugated momenta, and the Hamiltonian $(\ref{H})$ become operators:
\begin{equation}
     a \to \hat{a}, \qquad
     T \to \hat{T}, \qquad
    \hat{P}_{a} \to -i\frac{\partial}{\partial a}, \qquad \hat{P}_{T} \to -i\frac{\partial}{\partial T}, \qquad
     H \to \hat{H}.
\end{equation}
The equation that gives the quantum dynamics of the model is obtained when we require that the Hamiltonian operator annihilates the wavefunction $\Psi(\hat{a},\hat{T})$, which describes the physical properties of the Universe, at the quantum level. Thus, we have
\begin{equation} \label{WDW}
    \begin{matrix}
        \hat{H} \Psi(\hat{a},\hat{T}) = 0, \\
        \\
        \Big{(}\frac{1}{12} \frac{\partial^{2}}{\partial a^{2}} - 3a^{2} + \rho_v a^4-\frac{\sigma^{2}a^{4}}{(a^{3} + 1)^{2}}\Big{)} \Psi(a,\tau)  = -i\frac{\partial}{\partial \tau} \Psi(a,\tau),
    \end{matrix}
\end{equation}
where we made the change of variable $T=-\tau$. This equation above, given in the form of a time-dependent Schr\"{o}dinger equation, is called the Wheeler-DeWitt equation.

Assuming that one may write the solution to equation (\ref{WDW}) in the form $\Psi(a,\tau) = \psi(a)e^{-iE\tau}$, where $E$ is the energy of the radiation fluid. Using it in Eq. (\ref{WDW}), one finds,
\begin{equation}
    \frac{\partial^2}{\partial a^2} \psi(a) + 12[E - V_{eff}(a)] \psi(a)=0,
\label{WdW-2}
\end{equation}
\noindent with $V_{eff}(a)$ being given by Eq. (\ref{Veff}). Now, we solve Eq. (\ref{WdW-2}) using the WKB approximation. Once obtained, the WKB solution is used to calculate the tunneling probabilities through the potential barrier $V_{eff}(a)$.
 The tunneling probability, using the WKB solution, for a potential with two barriers is given in Ref. \cite{Bohm}. For our potential not centered at the origin, we have:
\begin{equation} \label{TP_WKB}
TP_{WKB} = \frac{4}{ \Big{(} \frac{\theta^{2}_{1} + \theta^{2}_{2}}{\theta_{1} \theta_{2}}\Big{)}^{2} \cos^{2}{(-\frac{J}{2} + \frac{\pi}{2})} + \Big{(} 4 \theta_{1} \theta_{2} + \frac{1}{4\theta_{1} \theta_{2}} \Big{)}^{2} \sin^{2}{(-\frac{J}{2} + \frac{\pi}{2})}} 
\end{equation}
where $\theta_{1}$ and $\theta_{2}$ represent the height and width of the first and second barriers, respectively, and $J$ represents the height and width of the well between the barriers. They are given by:

\begin{equation}
\begin{matrix}  \label{theta}
\theta_{1} = e^{{\int^{x_{2}}_{x_{1}}\hat{\kappa}(a)da}} \\
\\

\frac{J}{2} = e^{{\int^{x_{3}}_{x_{2}}\hat{K}(a)da}}\\
\\

\theta_{2} = e^{{\int^{x_{4}}_{x_{3}}\hat{\kappa}(a)da}}
\end{matrix}
\end{equation}
where $\hat{\kappa}(a) = \sqrt{12(V_{eff} - E)}$, $\hat{K}(a) = \sqrt{12 ( E - V_{eff})}$, $x_{1},x_{2}$ and $x_{3},x_{4}$ represent the points where a horizontal straight line of energy $E$ touches the effective potential (\ref{Veff}) from the left and right of the first barrier and the second barrier, respectively.


\section{Results} \label{seção3}

Once we have obtained equations $(\ref{TP_WKB})$ and $(\ref{theta})$, we can calculate the probability of the Universe emerging to the right of the double barrier potential given in Figure $(\ref{F1_Veff_2})$. By substituting equations $(\ref{theta})$ into $(\ref{TP_WKB})$, $TP_{WKB}$ becomes a function of three parameters: the energy $E$, $\rho_v$ the vacuum energy density, and the parameter $\sigma$ associated with the {\it ad hoc} potential $V_{ah}$ Eq. (\ref{ad hoc}). In this paper, we focus on studying the behavior of the tunneling probability as a function of these three parameters. We compute numerically $TP_{WKB}$ Eq. $(\ref{TP_WKB})$. 

The constant $\sigma$ associated with the {\it ad hoc} potential $V_{ah}$ Eq. (\ref{ad hoc}) is not directly constrained by the current observational data. This parameter controls the magnitude of $V_{ah}$. Therefore, in the absence of a fundamental theory that fixes its value, $\sigma$ must be interpreted as a phenomenological parameter. In the present analysis, its range is chosen so that the effective potential develops a double barrier structure, which is a necessary condition for the appearance of the resonant effect in tunneling. This choice is justified within the scope of the model, as it allows us to explore the physical implications of resonance phenomena in quantum cosmology.


We start by introducing the physical units in the typical expression $\theta_1$, Eq. (\ref{theta}) of the paper, contributing to the $TP_{WKB}$,
\begin{equation}
\label{theta1pu}
\theta_1 = e^{\left(\frac{1}{\hbar}\int_{x_1}^{x_2} \sqrt{12 m_p(V_{eff}-E)} l_p da\right)},
\end{equation}
where $m_p$ and $l_p$ are the Planck mass and length, respectively. Combining the quantities $\hbar$, $m_p$ and $l_p$, we may write Eq. 
(\ref{theta1pu}) as follows;
\begin{equation}
\label{theta1pu1}
\theta_1 = e^{\left(\int_{x_1}^{x_2} \sqrt{12 (V_{eff}-E)/E_p} da\right)},
\end{equation}
where $E_p$ is the Planck energy. Since we want to study the very early universe, let us consider that $V_{eff}$ and $E$ are of the order of the Planck energy $E_p = 1.9561 \times 10^9 J = 1.220890 \times 10^{19} GeV$. Now, all quantities present in the expression of $\theta_1$ have just numerical values and no physical units. After studying several double barrier potentials with different parameters values, we choose the following values for $\rho_v$ and $\sigma$:  $\rho_v=0.02925$ and $\sigma=-16.65$. They produce a good example of a double barrier potential. With this choice, the value of $\rho_v$ has almost the same order of magnitude of the Planck energy density. 
Therefore, in the expressions of $\theta_1$, $J/2$ and $\theta_2$ Eq. (\ref{theta}), we shall have the following $V_{eff}$,
\begin{equation}
\label{veffpu}
V_{eff}(a) = 3a^2 - 0.02925 a^4 + \frac{277.2225 a^4}{\left(a^3+1\right)^2}\quad.
\end{equation}
In the next subsections, we study how $TP_{WKB}$ behaves as a function of $E$, $\rho_v$ and $\sigma$.

\subsection{Tunneling Probability as a Function of Energy}

\begin{figure}[H] 
\centering
         \begin{minipage}[t]{0.6\textwidth}            
            \includegraphics[width=\linewidth]
            {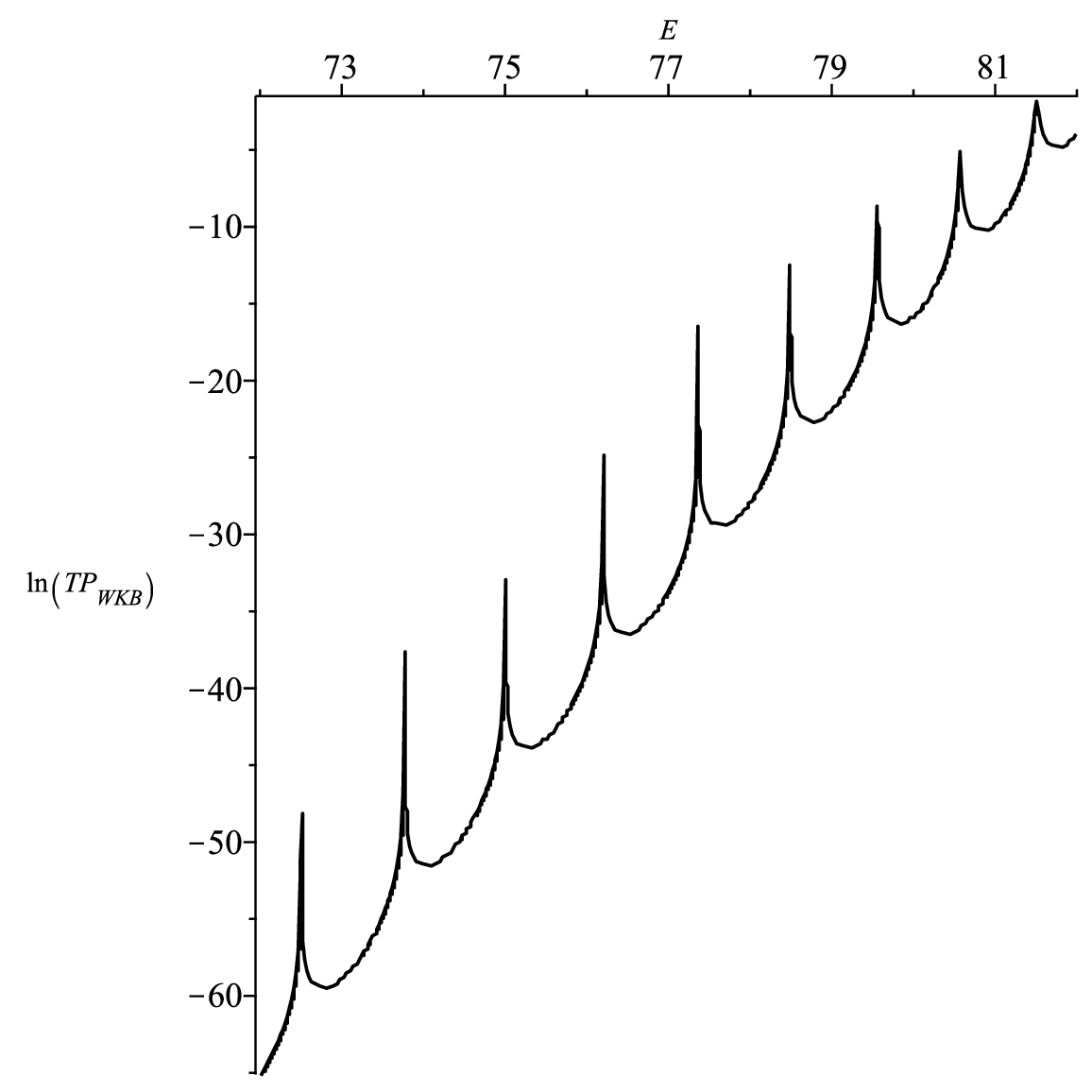}
        \end{minipage}
        \caption{Behavior of $TP_{WKB}$ as a function of energy $E$, in logarithmic scale, for 10001 energies. The variation starts at $E = 72$ and ends at $E = 82$, in intervals of $\Delta E = 0.001$, with $\sigma = -16.65$ and $\rho_v = 0.02925$.}
        
            \label{Variando E}
    \end{figure}

To study $TP_{WKB}$ as a function of the energy $E$, we fix the parameters values at $\sigma = -16.65$, $\rho_v=0.02925$ and leave the energy $E$ free to vary, in intervals of $\Delta E=0.001$, starting at $E=72$ and ending at $E=82$. We choose the maximum value of $E$ smaller than the height of the highest barrier $V_{eff}^{max} \approx 82.44$. Figure $\ref{Variando E}$ illustrates the behavior of $TP_{WKB}$ for $10001$ values of Energy.

    \FloatBarrier

When studying the tunneling probability for a potential with two barriers, the resonance phenomenon is expected in the literature \cite{Bohm, Merzbacher}. This phenomenon consists of the wave function undergoing multiple reflections when attempting to tunnel through the second barrier, causing the tunneling probability to increase considerably because the reflected waves have the same phase as the wave that passed through the first barrier, thus generating constructive interference. Due to multiple reflections between the barriers, when the potential shape is simple and symmetric (like square barriers), the tunneling probability graph as a function of $E$ is expected to present peaks where the probability reaches 100$\%$. On the other hand, for a more general two-barrier potential, this behavior does not occur \cite{Ressonância2,Ressonância1,Ressonância6,Ressonância7}. As we can see in Figure $\ref{Variando E}$, $V_{eff}$ Eq. (\ref{Veff}) is an example of a more general two-barrier potential. The graph $TP_{WKB}$ as a function of $E$ presents peaks where the intensity is much higher compared to neighboring points, but these peaks do not reach 100$\%$. Thus, the highest tunneling probability occurs for the energy $E=81.505$, which is less than the maximum barrier height $V_{eff}^{max} \approx 82.44$. Therefore, it is more likely that the universe is born with this resonant energy. Next, we study the behavior of $TP_{WKB}$ when we vary the other model parameters, for a fixed resonant energy value.

\subsection{Tunneling Probability as a Function of $\rho_v$}

\begin{figure}[H] 
\centering
         \begin{minipage}[t]{0.6\textwidth}            
            \includegraphics[width=\linewidth]{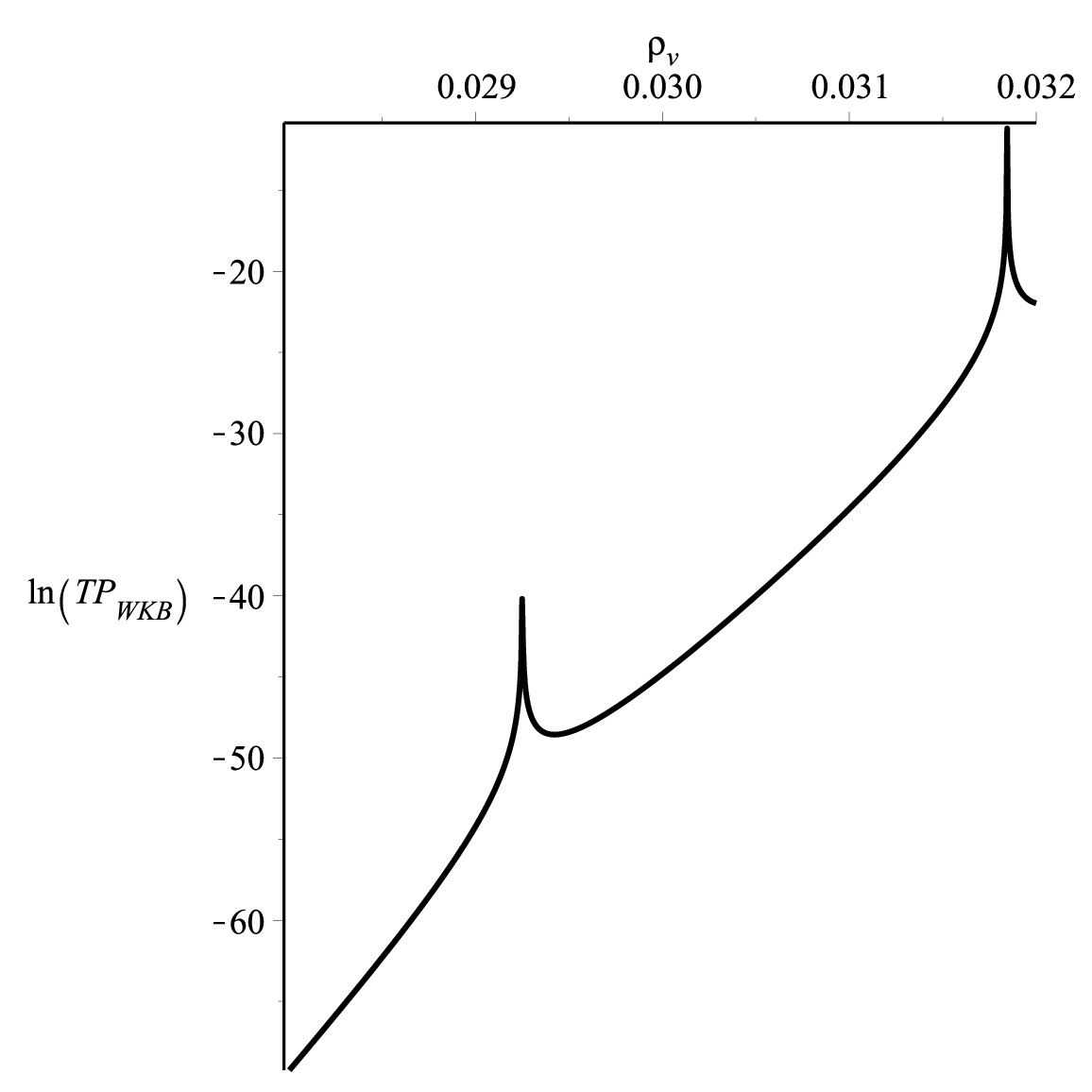}            
        \end{minipage}
        \caption{Behavior of $TP_{WKB}$ as a function of the parameter $\rho_{v}$, in logarithmic scale, for 800 values of $\rho_{v}$. The variation starts at $\rho_{v}=0.028000$ and ends at $\rho_{v}=0.032000$, in intervals of $\Delta \rho_{v}=0.000005$, with $E=73.76$ and $\sigma = -16.65$.}
            \label{Variando rho_v}
    \end{figure}

Now, consider the situation where $TP_{WKB}$ Eq. $(\ref{TP_WKB})$ is a function only of the parameter $\rho_{v}$. To do this, we fix the values of the other parameters at $E=73.76$, $\sigma = -16.65$ and leave the parameter $\rho_{v}$ free. In Figure $\ref{Variando rho_v}$, we see an example of the behavior of the tunneling probability as a function of $\rho_{v}$. This figure has $800$ values of $\rho_{v}$, where the parameter $\rho_{v}$ starts its variation at $\rho_{v}=0.028000$ and ends at $\rho_{v}=0.032000$, in intervals of $\Delta \rho_{v}=0.000005$. 
In Figure $\ref{Variando rho_v}$, we may see two resonant values of $TP_{WKB}$ for different values of $\rho_{v}$. Thus, the highest tunneling probability occurs for the second resonant value $\rho_{v} = 0.031845$, which is less than the maximum value of $\rho_{v}$ considered $\rho_{v_{max}} = 0.032000$. Therefore, it is more likely that the universe is born with this resonant value of $\rho_{v}$.

    \FloatBarrier



%
    
    \FloatBarrier

\subsection{Tunneling Probability as a Function of $\sigma$}

\begin{figure}[H] 
\centering
         \begin{minipage}[t]{0.6\textwidth}            
            \includegraphics[width=\linewidth]{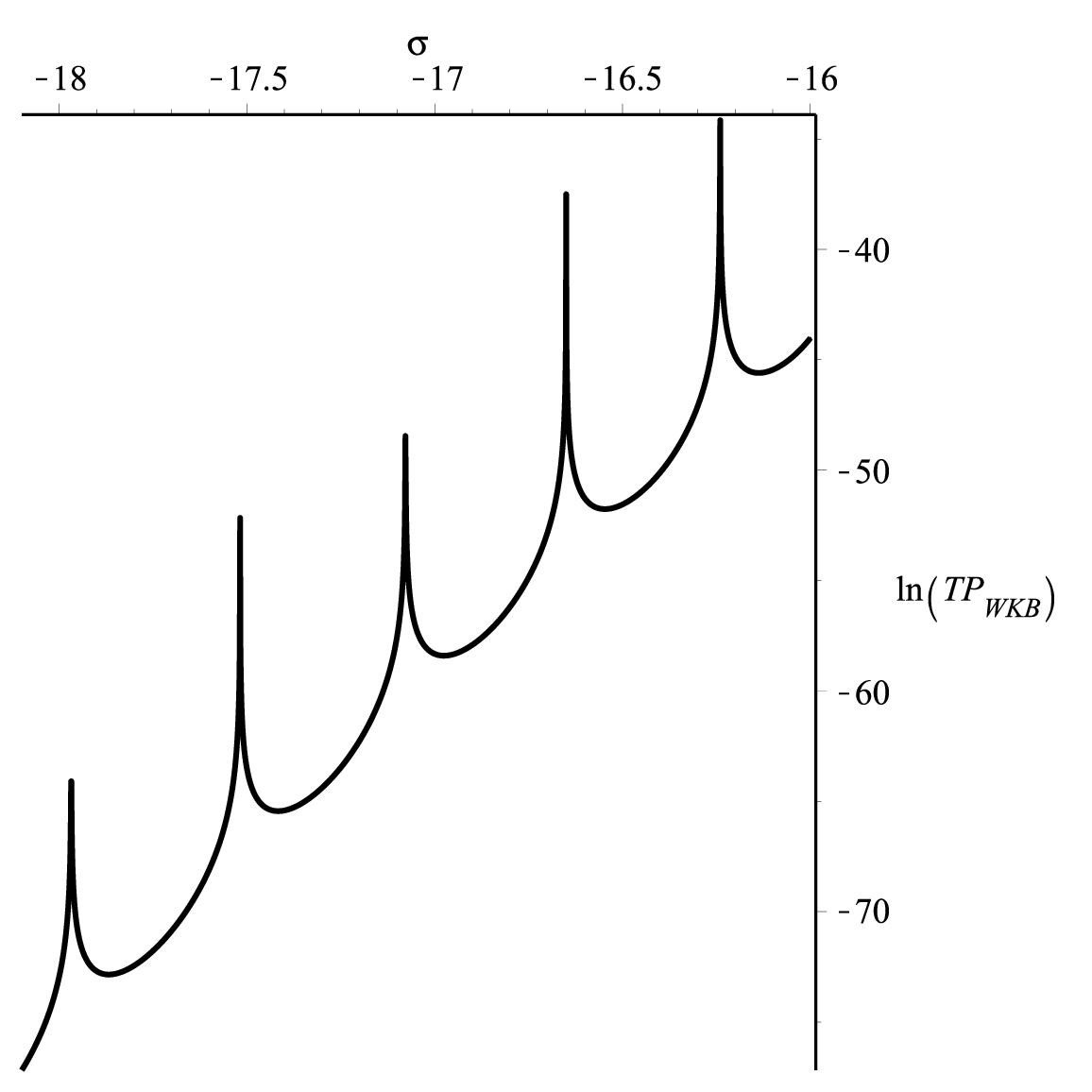}
            
        \end{minipage}
        \caption{Behavior of $TP_{WKB}$ as a function of the parameter $\sigma$, in logarithmic scale, for 2100 values of $\sigma$. The variation starts at $\sigma = -18.100$ and ends at $\sigma = -16.000$, in intervals of $\Delta \sigma=0.001$, with $E=73.76$ and $\rho_{v}=0.02925$.}
            \label{Variando C}
    \end{figure}

Now, we want to investigate the behavior of $TP_{WKB}$ as a function of $\sigma$. In this case, we fix the values of the other parameters at $E=73.76$ and $\rho_{v}=0.02925$ in Eq. $(\ref{TP_WKB})$. An example of the graph $TP_{WKB} \times \sigma$ can be seen in Figure \ref{Variando C}, where we have a graph with $2100$ values of $\sigma$, starting at $\sigma = -18.100$ and ending at $\sigma = -16.000$, in intervals of $\Delta \sigma=0.001$. Here, we can see the significant occurrence of resonances. Thus, the highest tunneling probability occurs for the resonant value $\sigma = -16.240$, which is less than the maximum value of $\sigma$ considered $\sigma_{max} = -16.000$. Therefore, it is more likely that the universe is born with this resonant value of $\sigma$.

    \FloatBarrier



%

  \FloatBarrier


\section{Conclusion} \label{conclusão}

In the present paper, we studied a Friedmann-Lemaître-Robertson-Walker (FLRW) quantum cosmology model with positively curved spatial sections. The matter content of the model was given by a radiation fluid, a constant vacuum energy, and an ad hoc potential. After writing the Hamiltonian of the model, we noticed that the effective potential ($V_{eff}$) depends on two parameters: $\rho_v$, the constant vacuum energy density, and $\sigma$, associated with the ad hoc potential. Depending on the values of these parameters $V_{eff}$ becomes a double barrier potential. We quantized the model and obtained the Wheeler-DeWitt equation. We solved that equation using the WKB approximation and computed the corresponding probability ($TP_{WKB}$) that the wavefunction of the universe tunnels through the double barrier potential $V_{eff}$. We studied how $TP_{WKB}$ behaves as a function of the parameters $\rho_v$, $\sigma$ and the radiation energy $E$. We noticed a significant occurrence of resonances in $TP_{WKB}$ when varying those parameters and $E$. It is a very interesting phenomenon because it may cause the universe to be born with selected values of those parameters and $E$. 

\section*{Acknowledgements}
\noindent D. L. Canedo thanks Coordenação de Aperfeiçoamento de Pessoal de Nível Superior (CAPES) and Universidade Federal de Juiz de Fora (UFJF) for his scholarships. G. Oliveira-Neto thanks FAPEMIG (APQ-06640-24) for partial financial support. G. A. Monerat thanks FAPERJ for partial financial support. G. A. Monerat and E. V. Corr\^{e}a Silva thank Universidade do Estado do Rio de Janeiro, UERJ, for the Proci\^{e}ncia grant.



\renewcommand{\theequation}{A.\arabic{equation}} 
\setcounter{equation}{0} 


\end{document}